\documentclass[twocolumn,prl,amsmath,amssymb,showkeys,showpacs]{revtex4}

\usepackage{dcolumn}
\usepackage{graphicx}
\usepackage{bm}

\begin{document}

\title{Quantum Vacuum Friction in Highly Magnetized Neutron Stars}

\author{Arnaud Dupays}
\affiliation{Laboratoire de Collisions, Agr\'egats, R\'eactivit\'e (UMR 5589, CNRS-Universit\'e de Toulouse, UPS), IRSAMC, Toulouse, France}

\author{Carlo Rizzo}
\affiliation{Laboratoire de Collisions, Agr\'egats, R\'eactivit\'e (UMR 5589, CNRS-Universit\'e de Toulouse, UPS), IRSAMC, Toulouse, France}

\author{Dimitar Bakalov}
\affiliation{INRNE, Bulgarian Academy of Sciences, Sofia, Bulgaria}

\author{Giovanni F. Bignami}
\affiliation{IUSS, Istituto Universitario di Studi Superiori, Pavia, Italy \\ 
INFN, Sezione di Pavia, Via A. Bassi 6, I-27100 Pavia, Italy}

\date{\today}

\begin{abstract}

In this letter we calculate the energy loss of highly magnetized neutron star due to friction with quantum 
vacuum, namely Quantum Vacuum Friction (QVF). Taking into account one-loop corrections in the effective 
Heisenberg-Euler Lagrangian of the light-light interaction, we derive an analytic expression for QVF allowing us to consider magnetic field at the surface of the star as high as $10^{11}\,$T. In the case of magnetars 
with high magnetic field above the QED critical field, we show that the energy loss by QVF dominates the 
energy loss process. This has important consequences, in particular on the inferred value of the magnetic 
field. This also indicates the need for independent measurements of magnetic field, energy loss rate, and of 
the braking index to fully characterize magnetars.

\end{abstract}

\maketitle

\section{Introduction}
Quantum Vacuum Friction (QVF) is a very fundamental phenomenon~\cite{Davies:04} 
related to the fact that quantum vacuum can be
regarded as a standard medium whith its own energy density and
electromagnetic properties. Well known effects due to static
vacuum properties are the Casimir effect~\cite{revCasimir} and the
vacuum magneto-electric optical properties~\cite{Rikken}.
Recently, dynamical effects due to vacuum viscosity have been
studied~\cite{Davies:04}, in particular photon radiation
stimulated by moving mirrors~\cite{Lambrecht:96} (i.e dynamical
Casimir effect) and friction between perfectly smooth surfaces moving relative to each other~\cite{Pendry:97,Feigel,Tiggelen}.

Considering a classical magnetic dipole moment rotating in
a standard medium with magneto-electric optical properties, relaxation
effects (at molecular level for example) produce retardation effects
between the induced magnetization of the medium and the rotation of the magnetic dipole moment. In this case, the
temporal delay between the medium's response and the inductor gives rise
to a classical dissipative frictional force.

In this letter, we investigate the friction effect between a non stationary magnetic dipole moment and its induced quantum
vacuum magnetic dipole moment. In this case the temporal delay between the vacuum's response and the inductor is due to the finite velocity
of light. We calculate the friction energy
loss rate in the case of highly magnetized neutron stars. Pulsars
are the more appropriate systems to look for such an effect.
Pulsars are fast rotating neutrons stars, with a very high
magnetic dipole moment tilted with respect to their rotational axis~\cite{ReviewPulsar}. 
Typically, their mass is of the order of that
of the Sun, and their radius of the order of $10$ km. The magnetic
field of neutron stars is typically of the order of $10^8-10^9$ T, 
while in the case of magnetars \cite{Kouveliotou:2003}
astrophysical observations seem to indicate that fields as high as
$10^{11}$ T exist on their surface~\cite{Ibrahim,Bignami,Ibrahim:04}. 
Spinning periods of neutron
stars range from tens of milliseconds for young stars to seconds,
increasing with star age. Because of their fast rotation retardation effects are significant. A rotating neutron star can thus be considered as a classical magnetic dipole moment weakly coupled to a bath, namely the quantum vacuum described by QED as electron-positron pairs due to quantum fluctuations. In a microscopic description, the energy is dissipated via the polarization process and through the annihilation of the polarized pairs which relax this excess of energy to the bath. This excess of energy corresponds to the frictionnal energy we calculate using a classical interaction between the star's dipole moment and the bath. 
We show that, for neutron star with a very high magnetic field, that exceeds the Quantum ElectroDynamics (QED) critical
field ($B_c= 4.4\,10^9\,$T), the energy loss is essentially due to
QVF, while the energy loss due to the classical process as star
rotating dipole radiation becomes negligible. This applies in
particular to magnetars which associate a strong magnetic field
and a few seconds spinning period. To infer the magnetic field of
such stars one cannot use the classical dipole energy loss
formula. Our results suggest that to characterize magnetars one
needs to measure the energy loss rate and the magnetic field
independently. Taking into account QVF, we calculate the braking index~\cite{manchester} 
and show that its measurement can provide a non-model dependent determination of the magnetic field on the surface of a neutron star.

\section{Energy loss rate}
To study the quantum vacuum magnetization in the presence of a
magnetic field, we start with the effective Heisenberg-Euler
Lagrangian ${\cal{L}}_{HE}$ of the light-light interaction~\cite{Euler}. Due to relativistic invariance, ${\cal{L}}_{HE}$ is a
function of the Lorentz invariants, $F$ and $G$, given by~\cite{landau}:
\begin{eqnarray}
\label{eq_fg}
F&=&\epsilon_0{\bf E}^2-\frac{{\bf B}^2}{\mu_0} \\
G&=&\sqrt{\frac{\epsilon_0}{\mu_0}}({\bf E}.{\bf B})
\end{eqnarray}
where $\epsilon_0$ is the vacuum permittivity and $\mu_0$ the vacuum
permeability. When one-loop corrections are included, ${\cal{L}}_{HE}$ can be
written as ${\cal{L}}_{HE}={\cal{L}}_{0}+{\cal{L}}_{1}$, where
${\cal{L}}_{0}=F/2$ is the usual Maxwell's term. Neglecting the Electric
field, $F=-{\bf B}^2/\mu_0$ and $G=0$. For ${\cal{L}}_{1}$ we use the analytic
expression derived by Heyl and Hernquist~\cite{Heyl0,Heyl}:
\begin{equation}
\label{eq_l1}
{\cal{L}}_{1}=\frac{e^2{\bf B}^2}{hc}X_0(\frac{B_c}{B})
\end{equation}
where $B_c=m_e^2c^3/e\hbar\simeq 4.4\,
10^{9}$ T is the critical field and
\begin{eqnarray}
\label{eq_X}
X_0(x)&=&4\int_0^{x/2-1}{\ln(\Gamma(v+1))dv} +
  \frac{1}{3}\ln{\left(\frac{1}{x}\right)} +2\ln{4\pi}\nonumber \\
&-&\left(4\ln{A}+\frac{5}{3}\ln{2}\right)-\left[\ln{4\pi}+1+\ln{\left(\frac{1}{x}\right)}\right]x
  \nonumber \\
&+&\left[\frac{3}{4}+\frac{1}{2}\ln{\left(\frac{2}{x}\right)}\right]x^2
\end{eqnarray}
with $\ln{A}=0.248\,754\,477$. Denoting by ${\bf M_{qv}}$ the induced magnetization, (i.e the quantum vacuum
  magnetic dipole moment per volume element), we have~\cite{Jackson}:
\begin{equation}
\label{eq_H}
{\bf H}=-2\frac{\partial {\cal{L}}_{HE}}{\partial{\bf
    B}}=\frac{{\bf B}}{\mu_0}-{\bf M_{qv}}
\end{equation}
Using the analytic form~(\ref{eq_l1}) of the Lagrangian, we can calculate
${\bf M_{qv}}$ to the first order of $\alpha=e^2/\hbar c\simeq 1/137$:
\begin{equation}
\label{eq_M}
{\bf M_{qv}}=\frac{\alpha B^2}{2\pi B_c^2 \mu_0}f_{qv}(B^2){\bf B}
\end{equation}
with
\begin{equation}
\label{eq_f}
f_{qv}(B^2)=\frac{1}{2}\left(\frac{B_c^4}{B^4}X_0^{(2)}\left(\frac{B_c}{B}\right)-\frac{B_c^3}{B^3}X_0^{(1)}\left(\frac{B_c}{B}\right)\right)
\end{equation}
where
\begin{equation}
\label{eq_Xn}
X_0^{(n)}=\frac{d^nX_0}{dx^n}
\end{equation}

\begin{figure}[h]
\includegraphics[width=8cm]{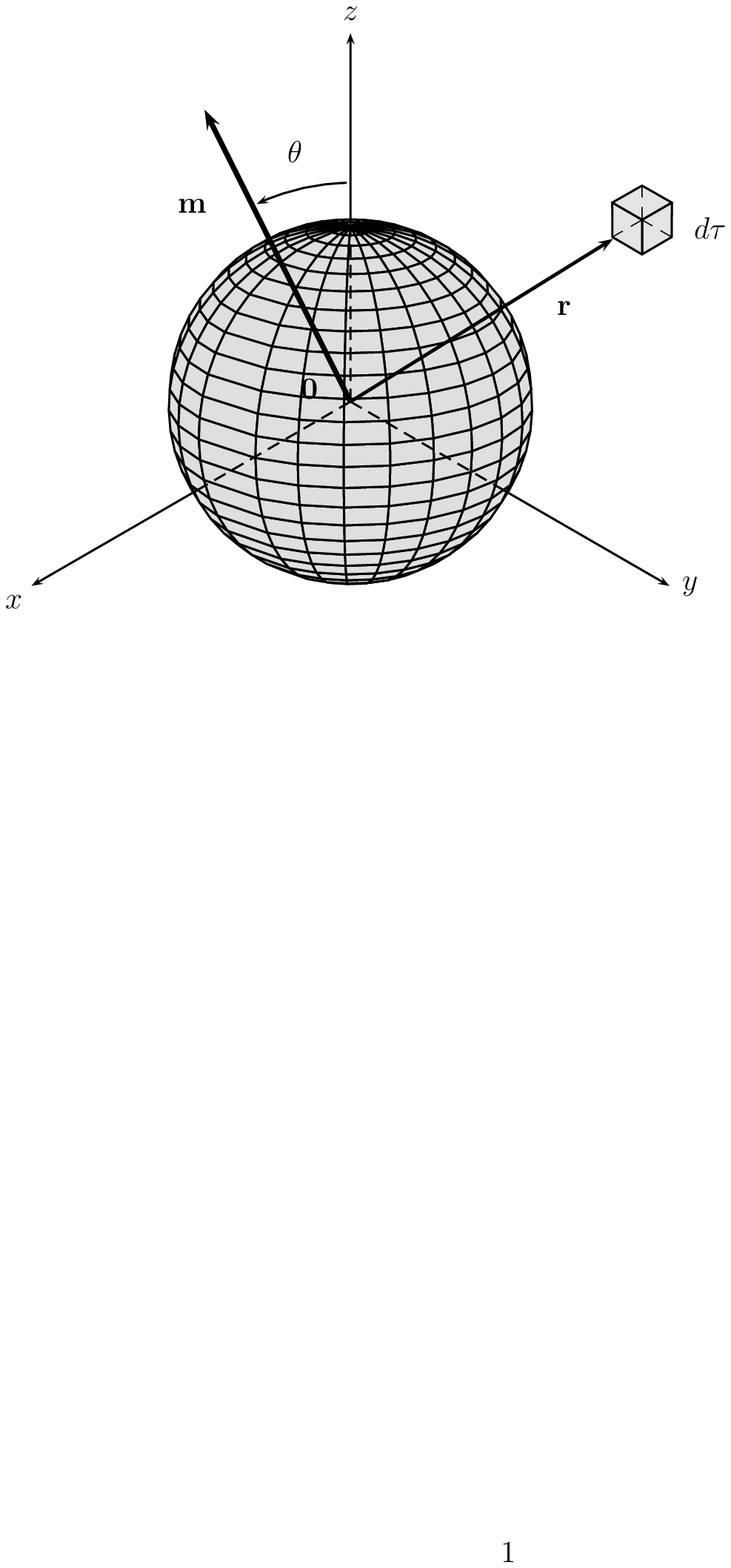}
\caption{\label{f_pulsar}{Rotating pulsar in vacuum. $\theta$ is the inclinaison of the star's magnetic dipole moment with respect to its rotation axis. $d\tau$ denotes a vacuum element of volume.}}
\end{figure}

Let us then now consider a neutron star rotating in vacuum (see fig.~\ref{f_pulsar}). We denote by ${\bf
  m}$ its magnetic dipole moment, $R$ its radius and $B_0$ the magnitude of the
  magnetic field at its surface ($B_0 \simeq \mu_0 m /4 \pi R^3$ where $m=\|{\bf m}\|$). We define a
  fixed frame $(x,y,z)$ with z-axis parallel to the rotation axis of
  this star. ($\theta,\varphi$) stand for the spherical polar angles
  of ${\bf m}$ in this fixed frame. If $p$ denotes the spinning period
  of the neutron star, $\varphi=\omega t$, with $\omega=2\pi/p$ and ${\bf
    m.u_z}=m\cos{\theta}$. At time $t$ the magnetic
  moment of the star produces a magnetic field ${\bf B(r,}t{\bf)}$. Let us
  denote by ${\bf r}$ the position vector of a vacuum element
  of volume $d\tau=r^2\sin{\beta}dr d\beta d\gamma$ where (r,$\beta,\gamma$) stand for the spherical coordinates
  of ${\bf r}$ in this fixed frame.. Since $\omega r/c \ll 1$ inside the region where quantum
  vacuum magnetization is important, the leading term within the dipolar
  magnetic field approximation is~\cite{Greiner} :
  \begin{equation}
    \label{eq_B}
    {\bf B(r,}t{\bf
    )}\simeq\left(\frac{\mu_0}{4\pi}\right)\left[\frac{3{\bf r}({\bf
        m}(t-r/c{\bf)}.{\bf r})}{r^5} -\frac{{\bf
        m}(t-r/c{\bf)}}{r^3}\right]
  \end{equation}
  In this expression, retardation effects have been taken into account with
  the argument $t-r/c$ in ${\bf m}$.
  According to eq.~(\ref{eq_M}), the induced quantum vacuum
  magnetic moment at ${\bf r}$ is given by :
  \begin{equation}
    \label{eq_dm}
    {\bf dm}_{qv}{\bf (r,}t{\bf)}=\frac{\alpha B^2({\bf r},t)}{2\pi B_c^2 \mu_0}f_{qv}(B^2({\bf r},t)){\bf B(r,}t{\bf)}d\tau
  \end{equation}
  At time $t+r/c$ the magnetic field ${\bf dB}_{qv}$ produced by ${\bf
  dm}_{qv}{\bf (r,}t{\bf)}$ at the center of the star is :
  \begin{eqnarray}
    \label{eq_dB}
    {\bf dB}_{qv}{\bf (0,}t+r/c{\bf
    )}&\simeq&\left(\frac{\mu_0}{4\pi}\right)\left[\frac{3{\bf r}({\bf
        dm}_{qv}{\bf (r,}t{\bf)}.{\bf r})}{r^5} \right. \nonumber \\
    &-&\left.\frac{{\bf
        dm}_{qv}{\bf (r,}t{\bf)}}{r^3}\right]
  \end{eqnarray}
This field interacts with the magnetic dipole moment of the
  star. At this stage, quantum vacuum can be regarded as a standard medium. Therefore, energy loss rate due to this friction is given by the classical formula~\cite{Jackson} :
  \begin{equation}
    \label{eq_fric}
    d\dot{E}_{qv}=-\left({\bf m(}t+r/c{\bf )}\times{\bf
    dB}_{qv}{\bf (0,}t+r/c{\bf )}\right)\omega.{\bf u_z}
  \end{equation}
  The total QVF energy loss rate is obtained by integrating eq.~(\ref{eq_fric}) over the space external to the star, and averaging it over time :
  \begin{equation}
    \label{eq_fric_tot}
    \dot{E}_{qv}=
    \int_{r=R}^{+\infty}\int_{\beta=0}^{\pi}\int_{\gamma=0}^{2\pi}
    \langle d\dot{E}_{qv} \rangle_t
  \end{equation}
$\dot{E}_{qv}$ can thus be calculated numericaly. Since retardation effects grow with $r$ while the star's magnetic field decreases significantly, the space region that mainly contributes to QVF is located around few star's radii and inside the light cylinder. Moreover, in the case of a
dipolar magnetic field, using eq.~(\ref{eq_f}), one can show that for $B_0 \leq10^{10}$ T,
$(B^2({\bf r},t)/B_c^2)f_{qv}(B^2({\bf r},t))\simeq (4B^2({\bf r},t)/45B_c^2)$
and we obtain the analytic expression  
\begin{equation}
  \label{eq_Bll}
  \dot{E}_{qv}\simeq
  \alpha\left(\frac{18\pi^2}{45}\right)\frac{\sin^2{\theta}}{B_c^2\mu_0c}\frac{B_0^4R^4}{p^2}
\end{equation}

\begin{figure}[h]
\includegraphics[width=8cm]{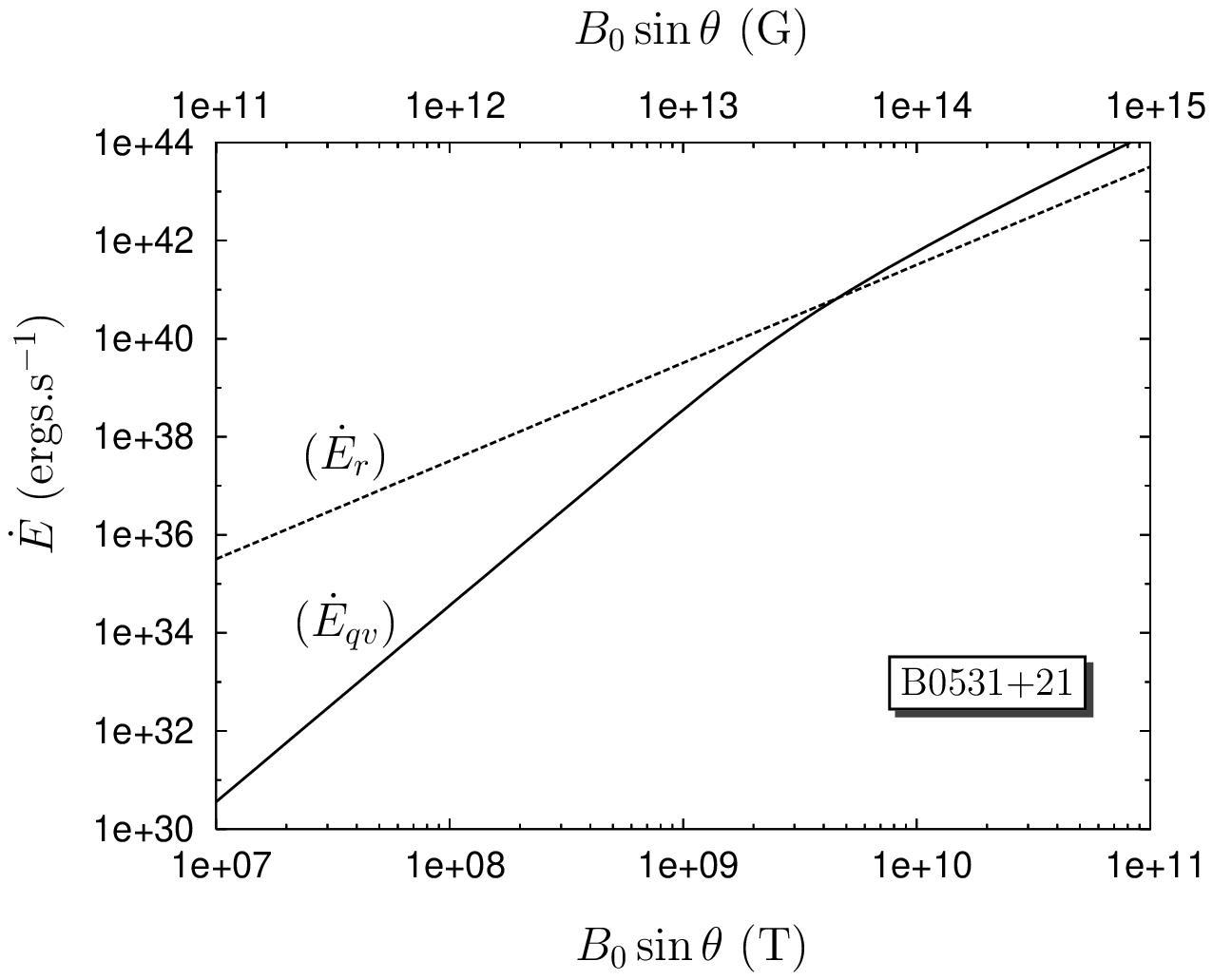}
\caption{\label{f_qvf1}{Quantum Vacuum Friction
    energy rate $\dot{E}_{qv}$ (full line) and classical matter radiative energy loss rate
    $\dot{E}_r$ (dotted line) as a function of $B_0\sin{\theta}$ for the Crab pulsar B0531+21.}}
\end{figure}

In fig.~\ref{f_qvf1}, we have plotted the function $\dot{E}_{qv}$ obtained numericaly from eq. (\ref{eq_fric_tot}) versus $B_0\sin{\theta}$ (full line), for the Crab pulsar with a spinning period $p=33,11\,ms$. We have also plotted the star rotation radiation rate
$\dot{E}_{r}$ versus $B_0\sin{\theta}$ given by classical dipole model~\cite{Pad} (since in this case, QED corrections are not relevant~\cite{Heyl}).

\begin{equation}
  \label{eq_EdotCla}
  \dot{E}_{r}= \left(\frac{128\pi^5}{3}\right)\frac{\sin^2{\theta}}{\mu_0c^3}\frac{B_0^2R^6}{p^4}
\end{equation}

Fig.~\ref{f_qvf1} clearly shows that for fields exceeding the QED
critical field the star energy loss is essentially due to QVF.

At this stage, it is important to stress that spinning period dependence is not
the same for the classical case and for the QVF. The ratio of classical to QVF losses decreases 
like $1/p^2$ for large $p$. QVF becomes more important for slowly rotating neutron stars. 

This result has important consequences. Actually, the value of the
magnetic field on the surface of a neutron star is inferred by the
energy loss of the star, derived by the measured value of the
spin-down rate, assuming that the energy loss of the star is given
by classical dipole model. In a case for which QVF should play an important role, 
the inferred value of the magnetic field should then be modified. Moreover, we see that the method used to get the value of the magnetic field, gives $B_0\sin{\theta}$ rather than $B_0$. One should thus assume the value of $\theta$ to get $B_0$ (typically $\theta$ is taken equal to $\pi/2$). In the following, taking advantage of the fact that this angular dependence is the same for $\dot{E}_{qv}$ and  $\dot{E}_{r}$, we show that the inferred value of the magnetic field at the surface of a pulsar can be derived by the measurement of the star's braking index.

\section{Braking index}
\begin{figure}[h]
\includegraphics[width=8cm]{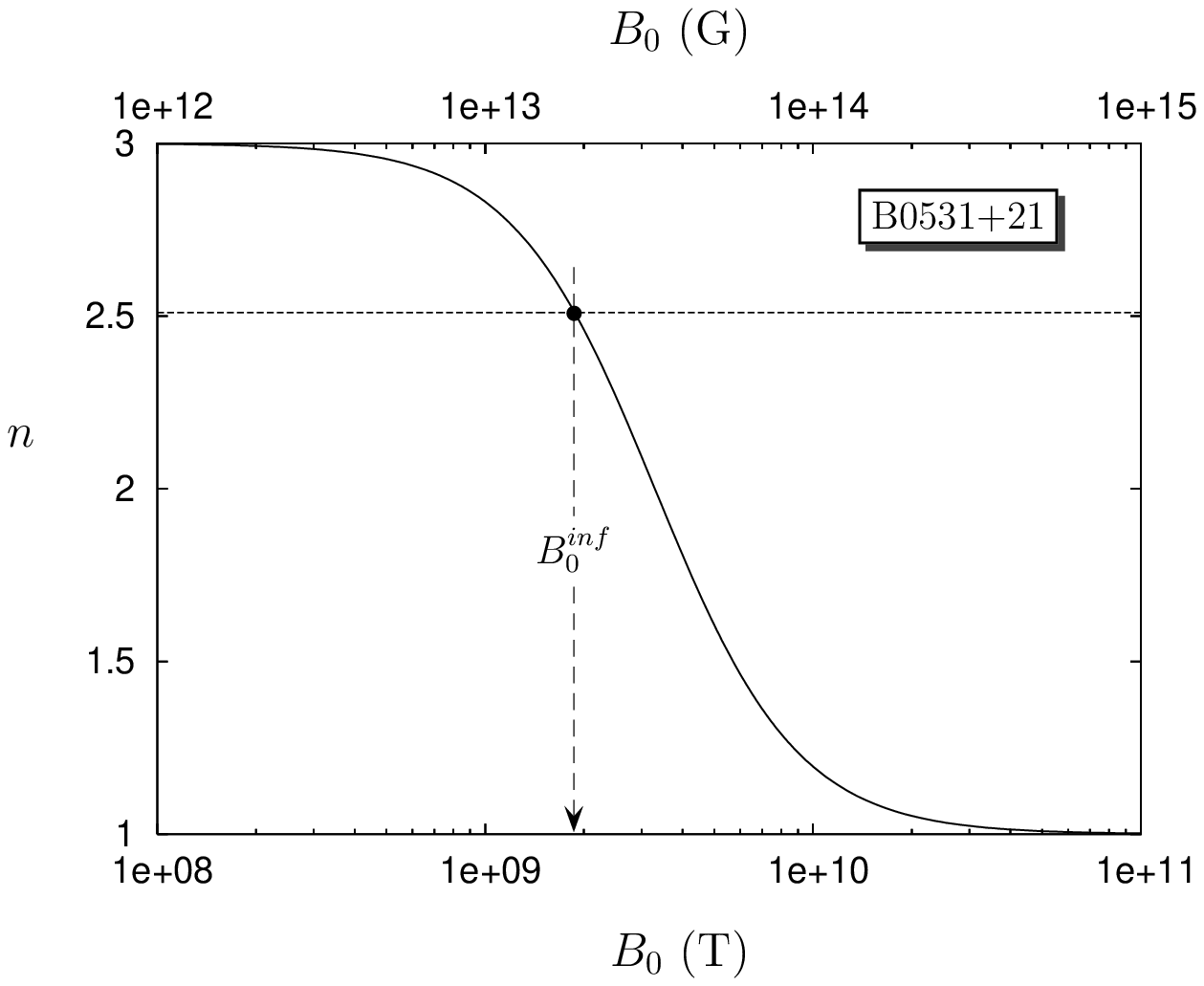}
\caption{\label{f_qvf2}{Derivation of the inferred value of the magnetic field at the surface of the Crab pulsar B0531+21 using the analytical expression given by eq.~(\ref{eq_braking_an}).}}
\end{figure}

The braking index $n$, is a fundamental parameter of pulsar electrodynamics describing the rate at which a magnetized neutron star loses rotational energy. This dimensionless quantity is given by 
\begin{equation}
\label{br_def}
n=\frac{\nu\ddot{\nu}}{\dot{\nu}^2}
\end{equation}
where $\nu$ is the spinning frequency of the pulsar and $\dot{\nu}$ (respectively $\ddot{\nu}$), denotes  the first (respectively the second) derivative of $\nu$ with respect to time. From this definition one can see that the braking index can be determined from pulsar timing measurements without any assumption concerning the star's structure and so,  provides a crucial information for our understanding of the physics underlying pulsar spin-down. 
From a theoretical point of view, the pulsar's rotational energy loss rate is given by
\begin{equation}
\label{e_rot_def}
\dot{E}=4\pi^2I\nu\dot{\nu}
\end{equation}   
where $I$ is the moment of inertia of the rotating star. Assuming a pure classical dipole energy loss mechanism, we obtain from eq.~(\ref{eq_EdotCla}) : 
\begin{equation}
\label{eq_nu_cla}
\dot{\nu}=-\frac{\dot{E}_r}{4\pi^2I\nu}=-K_r\nu^3
\end{equation}
where $K_r=\dot{E}_r/4\pi^2I\nu^4$. Taking the logarithmic derivative of this equation, we find $n=3$. So far, braking indices of only six pulsars have been precisely measured, all of which are remarkably smaller than the
value $n = 3$ expected for pure classical dipole radiation model (see table~\ref{tab_1}, second column). This result clearly shows that an additionnal energy loss mechanism should be included in the model to fit the data. Explanations of this discrepancy have already been suggested as for example the effect of the relativistic pulsar wind~\cite{Michel:69} or the fact that the pulsar's magnetism can not be modeled with a simple dipolar field~\cite{Melatos:97}. In this letter, we do not discuss the validity of such assumptions but we propose instead QVF as an additionnal energy loss mechanism which can provides a new coherente explanation for all the values of braking index measured so far. Taking into account QVF, eq.~(\ref{eq_nu_cla}) becomes 

\begin{equation}
\label{eq_nup}
\dot{\nu}=-K_r\nu^3-K_{qv}\nu
\end{equation}
where $K_{qv}=\dot{E}_{qv}/4\pi^2I\nu^2$. One can then derive the corresponding braking index 
\begin{eqnarray}
\label{eq_n}
n&=&\frac{3K_r\nu^3+K_{qv}\nu}{K_r\nu^3+K_{qv}\nu} \nonumber \\
&=&3-\frac{2}{1+\left(\frac{K_r}{K_{qv}}\right)\nu^2} 
\end{eqnarray}
which can be written as 
\begin{equation}
\label{n_lbo}
n=3-\frac{2}{1+\left(\frac{\dot{E}_r}{\dot{E}_{qv}}\right)}
\end{equation}
From this last expression together with eq.~(\ref{eq_Bll}) and eq.~(\ref{eq_EdotCla}) we see that the braking does not depend on $\theta$. More precisely we get
\begin{equation}
  \label{eq_braking_an}
  n=3-2\left[1+\left(\frac{320\pi^3}{3\alpha}\right)\frac{B_c^2 R^2}{B_0^2c^2}\nu^2\right]^{-1}
\end{equation}
In the case of magnetars ($B_0 \gg B_c$) we expect a value of $n \simeq 1$. In fig.~\ref{f_qvf2} we show how the inferred value of the magnetic field $B_0^{inf}$ can be obtained by matching the analytical expression given by eq.~(\ref{eq_braking_an}) with the measured value of a pulsar's braking index. $B_0^{inf}$ is thus coherent with $n$ and does not require any assumption on the pulsar's geometry.    

We have applied this method with the six braking indices wich have been measured with certainty. Results are presented table~\ref{tab_1} (third column) together with values of $B_0^{n=3}$ the magnetic field obtained with a pure classical dipolar model assuming $\theta=\pi/2$ and $n=3$ (last column). Depending of the case, the inferred value $B_0^{inf}$ can be higher or smaller than $B_{dipole}$. Nevertheless, it is interesting to notice that all the values found for the inferred magnetic field are smaller than the value of the QED critical field $B_c$, that is not the case with the classical dipolar model. This last result can provide a very important test to confirm QVF predictions. Because of the gamma-ray emitting electron-positron cascades which occur for $\dot{E}$ above the threshold value of $\dot{E}\simeq3\times10^{34} erg.s^{-1}$~\cite{Arons,Thompson:99}, magnetars, i.e pulsars with a magnetic field above the QED critical field (within the classical dipolar model) should not be gamma emitters. Detecting a gamma-ray emission from such a system, for example by the NASA GLAST mission due to launch in 2008, would provide an evidence in favor of QVF which leads to smaller values of the magnetic field for magnetars compatible with gamma-ray emission.

\begin{table}[h]
\begin{center}
\begin{tabular}{llccr}
\hline \\[-0.3cm]
Name & $n$ & $\nu$   & $B_0^{inf}$  & $B_0^{n=3}$  \\
          &        &  ($s^{-1}$)  &   ($10^{12}\,$G)  & ($10^{12}\,$G)  \\
\hline \\[-0.3cm]
\hline \\[-0.3cm]
J1846-0258  & 2.65(1) & 3.07 & 1.0 & 49     \\[0.15cm]
B0531+21  & 2.51(1) & 30.2  & 12.4 & 3.8    \\[0.15cm]
B1509-58  & 2.839(3) & 6.63  & 1.4 & 15     \\[0.15cm]
J1119-6127  & 2.91(5) & 2.45  & 0.4 & 42    \\[0.15cm]
B0540-69  & 2.140(9) & 19.8  & 12.4 & 5.1   \\[0.15cm]
B0833-45  & 1.4(2) & 11.2  & 16.1 & 3.4     \\[0.1cm]
\hline 
\end{tabular}
\end{center}
\caption{Spin and inferred magnetic field for pulsars with measured braking index $n$ taken from~\cite{Livingstone:2006,lyne_crab,Livingstone:2005a,Livingstone:2005b,Camilo:00,lyne_vela}.}
\label{tab_1}
\end{table}

\section{Conclusion} 
We have studied the energy loss due to friction
with quantum vacuum in the presence of high magnetized neutron
star, namely Quantum Vacuum Friction (QVF). In neutron
stars with magnetic fields of the order of $10^8-10^9\,$T, and
spinning period of less than a second, this effect is negligible
compare to classical energy loss dipolar process. In the case of
magnetars with high magnetic field, and spinning period of few
seconds, we have shown that the energy loss by QVF dominates the
energy loss process. This has important consequences, in
particular on the inferred value of the magnetic field. It also
indicates the need for independent measurements of magnetic field, energy loss
rate, and of the braking index to fully characterize magnetars. 
QVF should play as well an important role in the early stages of
neutron star formation and it should be taken into account in
models of magnetar (see e.g. \cite{Duncan:92}), also for what concerns its
spindown history. On the other hand, macroscopic violation of the linearity of Maxwell's equations predicted by QED~\cite{Rikken} has not yet been proved. Experimental search of vacuum birefringence is in progress (see~\cite{Battesti} and ref. within). Evidence for QVF in the neighbourhood of a neutron star would thus be the first confirmation of such fundamental QED prediction, while negative observation would be the first indication of limitations on the QED description of vacuum.

\acknowledgments
The authors thank P. Jones, S. Sarkar and D. A. Smith for helpful discussions.

\end{document}